\documentclass[11pt]{article}

\usepackage{amsmath,amsfonts,amssymb,amsthm,dsfont}
\usepackage[margin=1in]{geometry}
\usepackage[colorlinks]{hyperref}
\usepackage{mathtools}
\usepackage{amssymb}
\usepackage{bookmark}

\newcommand{\ket}[1]{|{#1}\rangle}
\newcommand{\bra}[1]{\langle{#1}|}

\begin{document}


\title{A detailed analysis of mathematics of entanglement in Non-Hermitian systems in real eigenvalue regime }

\author{Chetan Waghela \\ email \href{mailto:chetanwa@gmail.com}{chetanwa@gmail.com} 
}

\date{\today}
\maketitle

\begin{abstract}
Hamiltonian Mechanics works for conserved systems and Quantum Mechanics is given in Hamiltonian language. It is considered that complexifying the quantum Hamiltonian a balanced loss and gain model can be created. The usual mathematics of density operator formalism and entanglement is extrapolated to such systems and the consequences are studied. Namely, a complete formalism using Density operators is created for real eigenvalue regime of these Non-Hermitian systems and correct forms of Von-Neumann and Entanglement Entropy are created. The consequences are studied in this regime and depicted w.r.t recent papers by \cite{AKPati,Wang}.
\end{abstract}

\section{Notation}
\begin{description}

	\item[] h: Hermitian Hamiltonian
	\item[] H: Non Hermitian Hamiltonian 
	\item[] $\eta$: Metric for inner product of the state space of H
	\item[]$\pi$: square root of $\eta$ 
	\item[]$\mathcal{H}_h$ and $\mathcal{H}_H$ and $\mathcal{H}_{H^\dagger}$ are state spaces of h and H and $H^\dagger$ respectively 
	\item[]$A_h$ and $A_H$ are observables for h and H respectively.
	\item[]$\rho_h$ and $\rho_H$ are density matrices for h an H respectively. 
	\item[]$P_h$ and $P_H$ are projection operators for h and H respectively.
	\item[]V() is Von-Neumann entropy and E() is Entanglement entropy
	\item[]$\lambda$ and $\lambda'$ are eigenvalues of $H$ and $H^\dagger$
	\item[]$\Xi$ and $\xi$ are states for hermitian h, $\Psi$ and $\psi$ for Pseudo-Hermitian H and $\Phi$ and $\phi$ for Pseudo-Hermitian $H^\dagger$
\end{description}


\section{Introduction:}

Quantum Physicists has always known that a non-dissipating physical system can always be given by a Hermitian Hamiltonian which has all the ingredients of reality of eigenvalues (as they represent energy) and norm preserving time evolutions (as no particle or energy exchange is happening). In 1998 Bender et. al ,\cite{Bender1,Bender2,Bender3} studied eigenvalues of Non-Hermitian Hamiltonian, numerically. They found to their surprise that these Hamiltonians had a spectrum of Real eigenvalues in a certain parameter range. Moreover, these Hamiltonians were PT (Parity, time reversal)-Symmetric. However, the state space of such a system was not complete orthogonal (Hence, it couldn't form a Hilbert space), as it is needed to represent a non dissipative Quantum physical system ( which was until now always represented by Hermitian Hamiltonians). Several attempts were made by people to solve this difficulty,\cite{Bender4,Japaridze}. It was however shown by, Mostafazadeh,\cite{Mostafazadeh} that rather than focussing on PT-Symmetry one should focus on the Non-Hermitian property of these Hamiltonians, and he provided a 'general' framework which showed that all Non-Hermitian Hamiltonians in the real eigenvalue regime will show an exact behavior like systems represented by Hermitian Hamiltonians, by changing the metric of state space of these Hamiltonians. However, in the complex eigenvalue regime it can be seen that the system does not have norm preserving time evolutions even after changing metric and has complex eigenvalues, which is interpreted as onset of dissipation in this paper and ,\cite{Dekker1975,Dekker1981,Sergi} and many others. Now, the use of Non-Hermitian mathematical framework has been considered by ,\cite{Guo} where they have used it in classical optics by taking two waveguides, one of it loosing and other gaining, but overall the system is non-dissipating. They have specifically taken up the idea of symmetry breaking of non-hermitian systems, and used it in optical waveguide theory. We in this paper follow that the formalism of Pseudo-Hermiticity can be used to describe non-dissipating regime of a system whose dissipating nature depends on certain parameters ,\cite{Dekker1975,Dekker1981,Sergi} and it has been verified in the experiment that by going above certain threshold given by this theory  ( which ,\cite{Guo} called optical gain/loss coefficient) , a Non-Hermitian system starts dissipating. The outline of the paper is as follows, section 1,2,3,4,5 gives necessary and sufficient proofs for establishing the iso-spectrality between a Pseudo-Hermitian system and a Hermitian Hamiltonian system and addresses important subtleties creating confusion regarding this theory. Section 6 establishes the theory of measurement in these systems and Section 7,8 describes non-composite Pseduo-Hermitian systems in Density operator language. Section 8,9,10,11 describe Composite systems involving Pseudo-Hermitian systems and establish important Quantum measures and clarifications to .\cite{AKPati,Wang}

We will first of all clarify certain ideas related to this field and direct to Appendix for particular examples of the framework. There are certain ideas which need to be clarified using ' Geometry of projection of state space of a 2$\times$2  Non-Hermitian Quantum system  with real eigenvalues', which we have provided in Appendix. Also, hence forward we will designate Non-Dissipating systems in real eigenvalue regime as Pseudo-Hermitian.

Here we give minimal amount of proofs needed to establish this theory consistently.

We start by asking what is the relationship between eigenvalues of Non-Hermitian matrix and its hermitian conjugate.

\subsection{Theorem}
Consider an operator $H$ and $H^\dagger$ s.t( such that)  $H\neq H^\dagger$ and they obey an eigenvalue problem, then the eigenvalues of $H$ and $H^\dagger$, denoted by $\lambda$ and $\lambda'$ respectively are related as $\lambda=\lambda'^*$ ( where * denotes complex conjugation).
\subsection*{Proof:}
We know from given information that,
\begin{equation}
H\psi_n=\lambda_n\psi_n 
\end{equation}
and
\begin{equation} 
H^\dagger\phi_m=\lambda'_m\phi_m
\end{equation}
where $\psi$ and $\phi$ are eigenvectors of H and $H^\dagger$ respectively i.e
$\psi\in$ $\mathcal{V}_H$ and $\phi\in$ $\mathcal{V}_{H^\dagger}$ where $\mathcal{V}_H$ and $\mathcal{V}_{H^\dagger}$are vector spaces with inner product $\bra{}\ket{}$.Now,
taking conjugate transpose of (1)and then multiplying (1) by $\phi_n$ from right we have 
\begin{equation}
\psi{_n^\dagger}H^{\dagger}\phi_n=\lambda{_n^*}\psi{_n^\dagger}\phi_n
\end{equation}
using (2) we see that 
\begin{equation}
\psi{_n^\dagger}\phi_n\lambda'_n=\lambda{_n^*}\psi{_n^\dagger}\phi_n
\end{equation}
hence we can see that canceling $\psi{_n^\dagger}\phi_n$ that 
\begin{equation}
\lambda_n=\lambda'{_n^*}
\end{equation} 
Hence we can see that  when eigenvalues are real they are equal $\lambda_n=\lambda'_n$
\\
We know that matrices with same eigenvalues are similar \cite{Lang}, i.e. they are related by similarity transformation. We now check what will be the similar transformation between these Non-Hermitian matrices in real eigenvalue regime.

\subsection{Theorem}
In the real eigenvalue regime of H, H and $H^\dagger$ are related by $\eta$H$\eta{^{-1}}$=$H^\dagger$; s.t.
\begin{equation}
\eta=\sum_{i=1}^{n}\ket{\phi_i}\bra{\phi_i}
\end{equation} 
where n is the dimension of H, $\ket{\psi}\in\mathcal{V}_H$ and $\ket{\phi}\in\mathcal{V}_{H^\dagger}$. Note: $\mathcal{V}_H$ and $\mathcal{V}_{H^\dagger}$ are just state space of $H$ and $H^\dagger$ and they are not yet rendered into Hilbert space. 
\subsection*{Proof:}
We know $H^\dagger\ket{\phi_i}$=$\lambda'_i\ket{\phi_i}$
now conjugate transposing we get 
$$\bra{\phi_i}H=\lambda'{_i^*}\bra{\phi_i}$$
as proved earlier in real eigenvalue regime,
$$\lambda'_i=\lambda_i$$ 
we have $$\bra{\phi_i}H=\lambda'_i\bra{\phi_i}$$
multiplying $\ket{\phi_i}$ from left we have 
\begin{equation}
\ket{\phi_i}\bra{\phi_i}H=\lambda'_i\ket{\phi_i}\bra{\phi_i}
\end{equation}
also
multiplying $H^\dagger\ket{\phi_i}$=$\lambda'_i\ket{\phi_i}$ by $\bra{\phi_i}$ from right we have equating with (7) that
\begin{equation}
\ket{\phi_i}\bra{\phi_i}H=H^\dagger\ket{\phi_i}\bra{\phi_i}
\end{equation}
summing over i=1 to n gives
\begin{equation}
\eta{H}=H^\dagger\eta
\end{equation}
where $$\eta=\sum_{i=1}^{n}\ket{\phi_i}\bra{\phi_i}$$
Let us check the properties of the matrix $\eta$. 
Note: We have not yet proved that $\mathcal{H}_H$ i.e state space of H forms a Hilbert space (Complete Orthogonal vector space). We have not yet proved $\eta$ is invertible, such that we can always use the relation $\eta H\eta^{-1}$=$H^\dagger$ for finite dimensions.

\subsection{Theorem:}
If we have $H\neq{H^\dagger}$ and eigenvalues of $H$ are necessarily real, then ($\psi$,$\phi$) will be orthogonal to each other under $\bra{},\eta\ket{}$ as inner product and where, $\ket{\psi}\in\mathcal{V}_H$ and $\ket{\phi}\in\mathcal{V}_{H^\dagger}$
\subsection*{Proof:}
We know that $\eta$H$\eta{^{-1}}$=$H^\dagger$ in real eigenvalue regime $\therefore$
\begin{equation}
\eta\ket{\psi_m}=\ket{\phi_m}
\end{equation}
previous theorem gives
\begin{equation}
\eta\ket{\psi_m}=\sum_{i=1}^{n}\ket{\phi_i}\bra{\phi_i}\ket{\psi_m}
\end{equation}
Hence, by (10) we see that
\begin{equation}
\sum_{i=1}^{n}\ket{\phi_i}\bra{\phi_i}\ket{\psi_m}=\ket{\phi_m}
\end{equation}
therefore, $\bra{\phi_i}\ket{\psi_m}$ has to be $=\delta_{im}$ so that,
$$\sum_{i=1}^{n}\delta_{im}\ket{\phi_i}=\ket{\phi_m}$$
We can see that if we use the operator $\eta$ which we previously saw was the operator relating H and H$^\dagger$
by similarity transform, that 
\begin{equation}
\bra{\psi_k}\eta\ket{\psi_m}=\bra{\psi_k}\sum_{i=1}^{n}\ket{\phi_i}\bra{\phi_i}\ket{\psi_m}
\end{equation}
$=\delta_{ki}\delta_{im}=\delta_{km}$

Hence $\eta$ renders the state space of H, $\mathcal{V}_H$ to be now a complete orthonormal vector space, i.e a Hilbert space with sesquilinear form $\bra{}\eta\ket{}$ as the inner product. Let us now on denote it by $\mathcal{H}_H$. It can also be verified by example that Gram-Schmidt Orthogonalisation will not work for Non-Hermitian systems if it uses identity as metric.

\subsection*{Corollary1:}
We can easily see that $\eta$=$\eta^\dagger$:
$$\eta^\dagger=(\sum_{i=1}^{n}(\ket{\phi_i})(\ket{\phi_i})^\dagger)^\dagger$$
as (AB)$^\dagger$=B$^\dagger$A$^\dagger$, where A and B are Matrices.
Hence, we have $$\eta^\dagger=\sum_{i=1}^{n}{\ket{\phi_i}\bra{\phi_i}}=\eta$$

\subsection*{Corollary2:}
$\eta$ is always invertible see ,\cite{Lang}.
As $\eta$ has a full rank because, $\eta=\sum_{i=1}^{n}\ket{\phi_i}\bra{\phi_i}$ and all $\ket{\phi}$ are linearly independent of each other as they are eigenvectors of H$^\dagger$. Hence, the relation $\eta H\eta^{-1}=H^\dagger$ will always exist for finite dimensions.

\section{Note on Square roots of matrices in finite dimensions:}
Some emphasis has to be given to Square roots of matrices, because they will be useful to elaborate isospectrality.
A matrix has square root iff it is diagonalizable, and a Hermitian matrix is a always diagonalizable, hence square root of a Hermitian matrix always exists s.t. A$^\dagger$A=B, where A is square root of B. A matrix has many roots. A point to be noted is not all square roots of a Hermitian matrix be Hermitian (not properly emphasized even in ,\cite{Lang}), example:
$$A=\begin{pmatrix}
ae^{-i\theta}& b \\
b & ae^{i\theta}\\
\end{pmatrix}$$
We can see that 
$$A{^\dagger}A=\begin{pmatrix}
a^2+b^2& 2abe^{i\theta} \\
2abe^{-i\theta}& a^2+b^2\\
\end{pmatrix}=B$$
A$\neq$A$^\dagger$ but B=B$^\dagger$ where a,b$\in\mathds{R}$.
Also, it can be easily checked that as $\eta$ is always invertible it's root is also always invertible. This property has implications as we will see in Section 5.

\section{Pseudo-Unitarity:}
We already saw that  $\eta$H$\eta{^{-1}}$=$H^\dagger$; we define this property as Pseudo-Hermiticity, for hermitian H we see that $\eta$=$\mathds{1}$, identity matrix. Analogous to Unitary matrix U$^\dagger$=U$^{-1}$ similarly a matrix P we define Pseudo-Unitarity as  
\begin{equation}
\eta{P^{-1}}\eta^{-1}=P{^\dagger}
\end{equation} s.t.
\begin{equation}
\bra{\psi}P{^\dagger}\eta{P}\ket{\psi}=\bra{\psi}\eta{P^{-1}}\eta^{-1}\eta{P}\ket{\psi}=\bra{\psi}\eta\ket{\psi}
\end{equation}
Note: Unitary transformations preserve norm for Hermitian systems, then Pseudo-Unitary transformations preserve norm for Pseudo-Hermitian systems.

\section{Iso-spectrality of Pseudo-Hermitian Hamiltonian with Hermitian Hamiltonian:}
We saw that $\eta$ is Hermitian and hence it's root always exists. $\therefore$ We can always write $\eta=\pi{^\dagger}\pi$
Suppose, we define an operator 'h' by a similarity transform on Pseudo-Hermitian Operator 'H' then ,
$\pi$H$\pi^{-1}$=h
\subsection{Theorem:}
Prove that h is Hermitian.
\subsection*{Proof}
We know $\pi$H$\pi^{-1}$=h , then,
h$^{\dagger}$=($\pi^{-1}$)$^\dagger$H$^\dagger$$\pi^{\dagger}$, but H$^\dagger$=$\eta$H$\eta^{-1}$ so
\begin{equation}
h^\dagger=(\pi^{-1})^{\dagger}\pi{^\dagger}\pi{H}\pi^{-1}\pi{^{\dagger-1}}\pi^\dagger
\end{equation}
\begin{equation}
\therefore{h^\dagger}=\pi{H}\pi^{-1}=h
\end{equation}
As, $\pi$ always exists and hence h always will exist such that h is hermitian, H and h are isospectral partners because similarity transformation always preserves the eigenvalue of the operator. Moreover, we will see that the observables in Pseudo-Hermitian theory are different in form than that of Hermitian theory, but their values always exactly match that of isospectral hermitian partner. Hence, there will be no way to distinguish such systems in real eigenvalue regime.
\\
Moreover, we saw that for finite dimensions $\pi$ will always exist without exception as $\eta$ will always exist, and thus there will always exist a $h$ for corresponding $H$.

\subsection{Consequence on observables:}
In 'Conventional theory' an observable is defined as an operator which provides real expectation values, moreover adjointness of the operator is also important, usually in textbooks, Hermiticity and Self-Adjointness are synonymous with each other but it is not true. 
An, operator $A_h^@$ is called adjoint of $A_h$ if, 
$$\bra{\zeta}\ket{A_h\zeta}=\bra{A_h{^@}\zeta}\ket{\zeta}$$
Moreover, $A_h$ is called self-adjoint if $A_h=A_h{^@}$. So, as we see in Hermitian theory, H=H${^\dagger}$ and hence A is also self-adjoint in hermitian theory. In Pseudo-Hermitian theory as $\eta$H$\eta^{-1}$=H${^\dagger}$, H is called Pseudo-Hermitian, it is also called as Pseudo-self-adjoint 
$$\bra{\zeta}\eta\ket{H\zeta}=\bra{H^\dagger\zeta}\eta\ket{\zeta}$$.
In a similar fashion, all other observables in Pseudo-Hermitian theory are pseudo-self-adjoint i.e.
$$\eta{A_H}\eta^{-1}=A_H^{\dagger}$$
This, can be seen in the example of 2$\times$2 matrix given in Appendix, that Pauli matrices do not remain usual Pauli matrices in the Pseudo-Hermitian theory, otherwise we would get complex expectation values of spin. Now ,we will see that the states of Pseudo-Hermitian Hamiltonian have same expectation values of observables as given by states of it's Hermitian iso-spectral partners. Similar to H it can be seen that all observables are related to their Hermitian iso-spectral observables by transformation,
$$\pi{A_H}\pi^{-1}=A_h$$ where '$A_H$' is an observable for a Pseudo-Hermitian Hamiltonian system while '$A_h$' is an observable for a Hermitian Hamiltonian system, and $A_h=A_h^\dagger$.
As, we saw earlier, $\pi$H$\pi^{-1}$=h, where h=h${^\dagger}$ and $\pi{^\dagger}\pi$=$\eta$, $\eta$ being the metric of Hilbert space of H. So, $\pi\ket{\phi}=\ket{\xi}$, where $\ket{\psi}\in\mathcal{H}_H$ and $\ket{\xi}\in\mathcal{H}_h$.
$$\bra{\psi}\eta{A_H}\ket{\psi}=\bra{\xi}\pi{^{-1\dagger}}\pi{^\dagger}\pi\pi^{-1}{A_h}\pi\pi^{-1}\ket{\xi}=\bra{\xi}{A_h}\ket{\xi}$$ Hence, we can see that their expectation values match and there is no way to distinguish a Hermitian system from a system with 'Non-Hermitian Hamiltonian' in the real eigenvalue regime. However, as stated in Appendix the structure of the state space might seem different from that of it's Isospectral Hermitian partner, but it is not (see Appendix and Section 8 of ,\cite{Mostafazadeh}). \cite{AKPati} mentions that composite systems might provide a way to distinguish these systems, in this paper we show that even in these kind of systems we can find an isospectral hermitian hamiltonian, also we provide an information theoretic framework and show that 'Von-Neumann Entropy' of H and it's partner h provide same value provided the form of 'Von-Neumann Entropy' in Pseudo-Hermitian theory is changed in form just like the form of observables are changed. We also, show that Entanglement Entropy of composite systems of Hermitian$\otimes$Pseudo-Hermitian and Pseudo-Hermitian$\otimes$Pseudo-Hermitian will not change if they are transformed to their Isospectral partners under similarity transform. For this in later sections we will have to properly develop Pseudo-Hermiticity in Density operator language and try to leave no room for error.

\section{Usage of mathematics developed by Pseudo-Hermitian Theory:}
The method given in ,\cite{Dekker1975,Dekker1981,Sergi} to effectively describe a 'Dissipative system' involves, Complex Hamiltonian 
\begin{equation}
H_s=H_{s1}-i\Gamma(\theta)
\end{equation}
where, $H_s$ denotes the Hamiltonian for the system and $H_{s1}$ is Hermitian part (hence always has real eigenvalues) and $-i\Gamma(\theta)$ is the part through which we can control dissipation and is dependent on parameter $\theta$. Now, for the bath the Hamiltonian is given as, $H_b=H_{b1}+i\Gamma(\theta)$ s.t. $$H=H_s+H_b=H_{s1}+H_{b1}$$ 
where H is Hermitian and has real eigenvalues (hence overall it is not dissipating).

It can be seen that if only the system Hamiltonian is considered, then it is given as (18), and in the real eigenvalue regime as stated in previous two sections, the system acts like a Hermitian system. Note that it is not equivalent to making $\theta=0$ but even if $\theta\neq0$ the system acts as if it is not dissipating. It can be seen using the example $2\times{2}$ matrix given in Appendix. Also, it has been experimentally verified in .\cite{Guo}

\section{Measurements in Pseudo-Hermitian Theory}
In Hermitian theory we have two major types of measurements i.e. PVM and POVMs.\cite{McMahon}
PVMs: These are measurements where the observation does not destroy the observed state, repeated measurements provide same results. These are idealised measurements and are extremely rare to create. Mathematically, in Hermitian theory, these measurements are given by 'Projection' operators which have following two important properties, $P_h$.
$$P_h=P_h^{\dagger}$$ and $$P_h^2=P_h^{\dagger}P_h=P_h$$also $$\sum_{i=1}^{n}P_{h_i}=\mathds{1}$$ Now,
$$A_h=\sum_{i=1}^{n}c_i\ket{\xi_i}\bra{\xi_i}$$
where, $\ket{\xi_i}$ are basis of h; s.t. h=h$^\dagger$. If we define $$P_{h_i}=\ket{\xi_i}\bra{\xi_i}$$
then $$A_h=\sum_{i=1}^{n}c_iP_{hi}$$ all the above properties can be verified for $P_{hi}$.
Any state  $$\ket{\Xi}=\sum_{i=1}^{n}\ket{\xi_i}\bra{\xi_i}\ket{\Xi}=\sum_{i=1}^{n}c_i\ket{\xi_i}$$
where $c_i$ are the probability amplitudes for eigenvectors $\ket{\xi_i}$ of observable $A_h$ and they have complex values. The probability of obtaining particular ith state is . $$\mid{c_i^2}\mid=Pr(i)=\mid\bra{\xi_i}\ket{\Xi}{\mid^2}$$

After. the measurement $\Xi$ changes to $\Xi'$, if a measurement in such a type of scheme is performed again we get $\Xi''$ and it is same as $\Xi'$ \cite{McMahon}. These types of measurements are called 'PVM' or 'Projection operator valued Measurements'.
$$\ket{\Xi''}=\ket{\Xi}$$

\subsection{PVM prescription for Pseudo-Hermitian Theory:}
As the inner products are changed the outer products also need to be changed for consistency. Hence, the Projection operator for Pseudo-Hermitian Hamiltonian is given as 
\begin{equation}
P_{H_i}=\ket{\psi_i}\bra{\psi_i}\eta
\end{equation}
s.t. $$\ket{\Psi}=\sum_{i=1}^{n}\ket{\psi_i}\bra{\psi_i}\eta\ket{\Psi}=\sum_{i=1}^{n}c_i\ket{\psi_i}$$
$c_i\in\mathds{C}$ are probability amplitudes, and the probability of obtaining ith state, upon measurement, when a system is in state $\ket{\Psi}$ is
$$Pr(i)=\mid\bra{\psi_i}\eta\ket{\Psi}\mid^2=\bra{\Psi}\eta\ket{\psi_i}\bra{\psi_i}\eta\ket{\Psi}=\bra{\Psi}\eta{P_{H_i}}\ket{\Psi}=Tr({P_{H_i}}\ket{\Psi}\bra{\Psi}\eta)$$
Usually in Hermitian Theory we write $P^2=P$ but this statement is more subtle than this. The point is that for Complex matrices we reserve the definition $P_h^2$ for $$P_h^2=P_h^{\dagger}P_h$$ and in Hermitian theory we have $P_h^{\dagger}=P_h$ hence $$P_h^2=P_h^{\dagger}P_h=P_h{P_h}=P_h$$ as a fact of idempotency.
However if we follow the same definition of $P^2$ in Pseudo-Hermitian theory it creates ambiguities. But it can be seen that in Pseudo-Hermitian theory even if $P_H\neq{P_H}^\dagger$ $$P_H{P_H}=\ket{\psi_i}\bra{\psi_i}\eta\ket{\psi_j}\bra{\psi_j}\eta=P_H$$
and hence it can be seen that operating Projection operator $P_H$ twice does not affect the previous measurement result. Note that $P_H$ is a Pseudo-Hermitian operator. We will see what happens when repeated measurements affect the state after measurement.

\subsection{POVM prescription for Pseudo-Hermitian Theory:}
POVM  (Positive Operator Valued Measure) are more general type of measurements than PVMs. Unlike, PVM repeated measurements do not provide same measurement values, Ex: A photon falling on a photographic plate will have been altered drastically to be measured again. Mathematically, these are represented by more general 'Positive Semidefinite Operators', which  have following properties, $M_{h_m}\neq{M_{h_m}^\dagger}M_{h_m}$ always, m is just the index for denoting mth state. Post-measurement state is given by $\ket{\Xi'}=M_{h_m}\ket{\Xi}$ upto normalisation. but repeated measurement will not give same value. For, Pseudo-Hermitian theory, $M_{H_m}$ are selected such that they do not from $M_{H_m}=M_{H_m}M_{H_m}$ .

\section{Density Operator formalism for Pseudo-Hermitian theory:}
First, we develop a 'Density Operator Formalism' for Pseudo-Hermitian theory. This is done because this formalism provides more flexibility to talk about mixed states. Later we consider,composite systems and  quantify certain properties of these Quantum systems, like entanglement. Under this formalism we see that , there is no way that new phenomenon can emerge from composite systems of Pseudo-Hermitian$\otimes$Hermitian systems(denoted H$\otimes$h) and Pseudo-Hermitian$\otimes$Pseudo-Hermitian (denoted H${_1}$ $\otimes$H${_2}$) systems. Also, it is shown that the form of 'Entanglement entropy'(E()) as well as of 'Von Neumann entropy' (V()) has to be changed and under this new form it can be shown (for both cases stated above) that the values E() of  an entangled state in H and of an isospectral entangled state in h will be same, hence the entanglement entropy does not reduce as stated in ,\cite{AKPati} due to isospectrality. Also, it is easy to prove that rate of entanglement of both of these types of systems will be same.

\subsection{Pure states in Pseudo-Hermitian Theory:}
Let us usually denote Hermitian systems by h and it's eigenvectors as $\ket{\xi}$ which form an orthogonal basis for $\mathcal{H}_h$ Hilbert space for h.
$$\rho_{h_p}=\ket{\Xi}\bra{\Xi}$$. $\rho_{h_p}$ is the density operator for pure states of h. See ,\cite{McMahon}. Where, $\ket{\Xi_z}=b_{z_1}\ket{\xi_1}+b_{z_2}\ket{\xi_2}+....+b_{z_n}\ket{\xi_n}$, where 'z' denotes state of the system and second index denotes the basis vector of the system being considered, also $b_{z_1}^2+b_{z_2}^2+.....+b_{z_n}^2=1$, are probability amplitudes. Following relations always hold for such systems, $\rho_{h_p}=\rho_{h_p}^\dagger$ and Tr$(\rho_{h_p})=1$, and Tr$(\rho_{h_p}^2)=1$, because $\rho_{h_p}^2=\rho_{h_p}$.
Also, expectation values of any observable $A_h$ is given by $<A_h>$=Tr$({A_h}\rho_{h_p})$ All these properties can be checked taking a state as $\ket{\Xi_z}=\sqrt{3/4}\ket{0}+\sqrt{1/4}\ket{1}$.
For Pseudo-Hermitian system H, with eigenvectors $\ket{\psi}$ and having a hilbert space $\mathcal{H}_H$ with metric $\eta$, s.t. $\eta=\sum_{i=1}^{n}\ket{\phi_i}\bra{\phi_i}$ where $\ket{\phi}\in\mathcal{H}_{H^\dagger}$, n being dimension of H. 

Analogously for pure states of Pseudo-Hermitian systems, we define
\begin{equation}
\rho_{H_p}=\ket{\Psi_z}\bra{\Psi_z}\eta
\end{equation}
sticking to this definition, we can see that $\rho_{H_p}^\dagger=\eta\ket{\Psi_z}\bra{\Psi_z}$, where m denotes the state of system. Note, $\rho_{H_p}^\dagger\neq\rho_{H_p}$. but it can be checked that 
\begin{equation}
\rho_{H_p}^\dagger=\eta\rho_{H_p}\eta^{-1}
\end{equation}
i.e. it is Pseudo-Hermitian. Tr$(\rho_{H_p})$=$\sum_{i=1}^{n}\mid{c_i}\mid^2\bra{\psi_i}\eta\ket{\psi_i}$=1 this happens because $\sum_{i=1}^{n}\mid{c_i}\mid^2$=1 . Also, $\rho_{H_p}^\dagger\rho_{H_p}\neq\rho_{H_p}$,but
\begin{equation}
\eta^{-1}\rho_{H_p}^\dagger\eta\rho_{H_p}=\rho_{H_p}\rho_{H_p}=\rho_{H_p}
\end{equation}
The expectation value of an observable $A_H$ is given as 
\begin{equation}
<A_H>=Tr(A_H\rho_{H_P})=\bra{\Psi_m}\eta{A_H}\ket{\Psi_m}
\end{equation}
Now, using the Iso-spectrality property we can see that, $\ket{\Psi_z}=\pi^{-1}\ket{\Xi_z}$ where $\pi$ is the similarity transformation operator between Pseudo-Hermitian hamiltonian H to it's iso-spectral Hermitian Hamiltonian h.
\begin{equation}
\rho_{H_p}=\ket{\Psi_z}\bra{\Psi_z}\eta=\pi^{-1}\ket{\Xi_z}\bra{\Xi_z}(\pi^{-1})^\dagger\eta=\pi^{-1}\rho_{h_p}\pi
\end{equation} and it can also be checked that, using cyclic property of trace, $<A_H>=<A_h>$ 
\begin{equation}
<A_H>=Tr(A_H\rho_{H_p})=Tr(\pi^{-1}{A_h}\pi\pi^{-1}\rho_{h_p}\pi)=Tr(A_h\rho_{h_p})=<A_h>
\end{equation}
H and h are indistinguishable.

\subsection{Statistical mixtures containing Pseudo-Hermitian systems:}
There are two cases which arise for statistical mixtures, (Pseudo-Hermitian and Pseudo-Hermitian) and (Pseudo-Hermitian and Hermitian) mixtures.
Firstly, revising Hermitian systems theory, state of mixtures cannot be written as $\ket{\Psi}=c_1\ket{\psi}+c_2\ket{\psi}...$. They are formed due to incomplete information. Example: A screen which is impinged upon by 70\% of vertically polarised light from one source and 30\% from other being horizontally polarised. They  can only be written in Density Operator form. $\rho_{h_m}=\sum_{i=1}^{l}p_i\ket{\Xi_i}\bra{\Xi_i}=\sum_{i=1}^{l}\rho_{ih_m}$ , where p$_i$ denote probability of finding a certain state forming the mixture and $\sum_{i=1}^{l}p_i=1$, 'l' is the number of systems forming the mixture.(ex:$\rho_{h_m}=(3/4)\ket{0}\bra{0}+(1/4)\ket{1}\bra{1}$). 75\% are impinged upon from system 1 as $\ket{0}$ and other 30\% from system 2 as $\ket{1}$. Tr$(\rho_{h_m})=\sum_{i=1}^{l}p_i=1$ and Tr$(\rho_{h_m}^2)=\sum_{i=1}^{l}p_i^2<1$.

\subsubsection{Case 1:Statistical mixtures of only Pseudo-Hermitian Hamiltonians}
Density operator in such conditions is given as $$\rho_{H_m}=\sum_{i=1}^{l}p_i\ket{\Psi_i}\bra{\Psi_i}\eta_i=\sum_{i=1}^{l}\rho_{iH_m}$$ where, $\eta_i$ are the metric for corresponding systems of which mixture is made of,  $\sum_{i=1}^{l}p_i=1$ where p denote the percentage of the individual systems which make the mixture, l is the number of individual systems making up the mixture and $\rho_{iH_m}$ be individual density operators. Analogous to Hermitian case, $$Tr(\rho_{H_m})=\sum_{i=1}^{l}{p_i}\bra{\Psi_i}\eta\ket{\Psi_i}=\sum_{i=1}^{l}{p_i}=1$$.
Due to the Iso-Spectrality it can be seen that individual systems $$\rho_{iH_m}=\pi_i\rho_{ih_m}\pi_i^{-1}$$and we can write, $$A_{ih}=\pi_iA_{iH}\pi_i^{-1}$$ and we get the expectation values, by substituting previous results
\begin{equation}
<A_H>=Tr(\sum_{i=1}^{l}A_{iH}\rho_{iH_m})=Tr(\sum_{i=1}^{l}A_h\rho_{ih_m})=<A_h>
\end{equation}

\subsubsection{Case 2:Statistical Mixture of Pseudo-Hermitian and Hermitian Hamiltonians}
Out of total 'l' mixtures consider a mixture of 'e' Pseudo-Hermitian systems and 'l-e' Hermitian systems. Then, the density operator is defined as, 
\begin{equation}
\rho_{H_m}=\sum_{i=1}^{e}{p_i}\ket{\Psi_i}\bra{\Psi_i}{\eta_i}+\sum_{j=e}^{l}{p_j}\ket{\Xi_j}\bra{\Xi_j}=\sum_{i=1}^{e}\rho_{iH_m}+\sum_{j=e}^{l}\rho_{jh_m}
\end{equation}
where $\ket{\Psi_i}\in\mathcal{H}_{H_i}$ and $\ket{\Xi_j}\in\mathcal{H}_{h_j}$ and $\eta_i$ are metric for corresponding Hilbert spaces of Pseudo-Hermitian systems. Here, as in previous case we define $\rho_{iH_m}$ and $\rho_{jh_m}$, also as usual notation H correspond to Non-Hermitian and h correspond to Hermitian, also
$$\sum_{i=1}^{e}p_i+\sum_{j=e}^{l}p_j=1$$ 
It can also be easily checked that, Tr$(\rho_{H_m})$=1. Also as we saw in previous section, $A_{iH}={\pi_i}A_{ih}{\pi_i^{-1}}$ and and the relation $\pi_i\ket{\Psi_i}=\ket{\Xi_i}$, where these new $\ket{\Xi_i}$ belong to Hermitian isospectral system of H$_i$s. Now, we get using all these results that,
\begin{equation}
<A_H>=\sum_{i=1}^{e}{p_i}\bra{\Xi_i}(\pi^{-1})^\dagger\pi{^\dagger}\pi{\pi^{-1}}A_{ih}\pi{\pi^{-1}}\ket{\Xi_i}+\sum_{j=e}^{l}p_j\bra{\Xi_j}A_h\ket{\Xi_j}
\end{equation}
$$=\sum_{i=1}^{l}p_i\bra{\Xi_i}A_h\ket{\Xi_i}$$

\begin{equation}
\therefore <A_H>=<A_h>
\end{equation}

It can be seen in all the scenarios of a non-composite system,the open system in real eigenvalue regime acts like a Hermitian system(Non-Dissipating system).

\section{Von-Neumann Entropy: }
As we changed the form of observables in a hermitian setting from $A_h$ to $A_H$, by relation $\pi{A_H}\pi^{-1}=A_h$. An example is properly elaborated in the Appendix that the representation of Pauli Matrices need to be changed to properly get real expectation values of spin. It is a simple fact that due to isospectrality the values of entropy of H and h will be same because the Von-Neumann Entropy is defined as 
$$-\sum_{i=1}^{n}\lambda_i\log{\lambda_i}$$, where $\lambda_i$ are the eigenvalues of $\rho_H$ and $\rho_h$ both. This happens due to similarity equivalence between them, and similarity transformations preserve eigenvalues. This is contrast to ideas taken up by ,\cite{AKPati,Wang} where they have ignored that the two equivalent formulae for Von-Neumann entropy do not follow in the same way as in Pseudo-Hermitian theory. This can be proved in Density matrix form using the iso-spectrality condition. In Hermitian setting, another way of quantifying Von-Neumann entropy is given as $-Tr(\rho_h\log{\rho_h})$ where logarithm is to the base 2. For, Pseudo-Hermitian Hamiltonians however the proper form of Von-Neumann Entropy should be given as below, so that there is consistency in definition of Von-Neumann entropy
\begin{equation}
V(\rho_H)=-Tr(\pi\rho_H\pi^{-1}\log{(\pi\rho_H\pi^{-1})})
\end{equation}
Then it can be easily seen that due to the transformation. $\pi\rho_H\pi^{-1}=\rho_h$, that
\begin{equation}
V(\rho_H)=V(\rho_h)
\end{equation}

\section{Composite systems in Hermitian case}
Let us first revise Composite systems in Hermitian theory. Usually if two systems are to be represented simultaneously, then these systems are to be represented using tensor product $\otimes$ of the matrices representing them. Let us consider for simplicity of case of bipartite (i.e 2 systems) composite system, which can be generalized to many more systems (multipartite) by generalization. As we are for revision considering bipartite hermitian composite system, we denote the hamiltonians of these systems as, $h_1$ and $h_2$, which form a composite system $h$ denoted by, $h_1{\otimes}h_2$. We will denote the dimensions of these systems by E and F. Any state residing in the Hilbert space of such a system is given by, 
\begin{equation}
\ket{\Xi}=\sum_{i,j=1}^{E,F}c_{ij}\ket{\xi_i}_1\otimes\ket{\xi_j}_2
\end{equation} 
where $\ket{\xi}_1\in\mathcal{H}_{h1}$ and  $\ket{\xi}_2\in\mathcal{H}_{h2}$ and $\mathcal{H}_h=\mathcal{H}_{h1}\otimes\mathcal{H}_{h2}$ and $c_{ij}\in{C}$ are probability amplitudes as in $\sqrt{1/2}\ket{0}+\sqrt{1/2}\ket{1}$.

We will particularly focus on what happens to inner product metric for such systems, it is not always explicitly emphasized that metric of inner product of a bipartite systems is $\mathds{1}=\mathds{1}_1\otimes\mathds{1}_2$. These metrics are such that $_1\bra{\xi_i}\mathds{1}_1\ket{\xi_j}_1=\delta_{ij}$ and $_2\bra{\xi_i}\mathds{1}_2\ket{\xi_j}_2=\delta_{ij}$.

$\ket{\Xi}$ is called 'seperable' if it can be decomposed as,

$$\ket{\Xi}=\sum_{i,j=1}^{E,F}c_{ij}\ket{\xi_i}_1\otimes\ket{\xi_j}_2=(\sum_{i=1}^{E}c_{i}\ket{\xi_i}_1)\otimes(\sum_{j=1}^{F}c_{j}\ket{\xi_j}_2)$$. If it cannot be decomposed as earlier then it is called 'Entangled'.

A few important results while using tensor products are as such, 

\begin{equation}
({A_1}\otimes{B_2})({C_1}\otimes{D_2})={A_1}{C_1}\otimes{B_2}{D_2}$$and$$
(A\otimes{B})^{\dagger}={A^\dagger}\otimes{B^\dagger}
$$and$$
(A\otimes{B})^{-1}=A^{-1}\otimes{B^{-1}} 
$$and$$
{A_1\otimes{B_2}}\neq{B_2\otimes{A_1}}
\end{equation}

A,B,C,D can be any column, row , square matrices.

Therefore, \begin{equation}
\bra{\Xi}\ket{\Xi}=(\sum_{i,j=1}^{E,F}c_{ij}^*\bra{\xi_i}_1\otimes\bra{\xi_j}_2)(\mathds{1}_1\otimes\mathds{1}_2)(\sum_{i',j'=1}^{E,F}c_{i'j'}\ket{\xi_{i'}}_1\otimes\ket{\xi_{j'}}_2)=\sum_{i,j=1}^{E,F}\mid{c_{i,j}}\mid^2
\end{equation}
$_1\bra{\xi_i}\mathds{1}_1\ket{\xi_j}_1=\delta_{ij}$and $_2\bra{\xi_i}\mathds{1}_2\ket{\xi_j}_2=\delta_{ij}$ and using the properties stated above.

\subsection{Density Matrix Formulation of Composite systems in hermitian case:}
For simplicity, we will consider henceforward, that we are considering systems in pure states. Density matrix of a composite system is given as,$$\rho_{h_p}=\ket{\Xi}\bra{\Xi}(\mathds{1}_1\otimes\mathds{1}_2)$$ where, $\mathds{1}_1\otimes\mathds{1}_2=\mathds{1}$.

and $$Tr(\rho_{h_p})=\bra{\Xi}(\mathds{1}_1\otimes\mathds{1}_2)\ket{\Xi}=\sum_{i.j=1}^{E,F}={\mid}{c_{i,j}}{\mid^2}$$  

Von Neumann entropy is given explicitly as
\begin{equation}
V(\rho_{h_p})=-Tr(\rho_{h_p}\log_2(\rho_{h_p}))=-Tr(\ket{\Xi}\bra{\Xi}\log_2\ket{\Xi}\bra{\Xi})=-(\sum_{i=1}^{S}\kappa_i\log_2\kappa_i)
\end{equation}
where $h=h_1\otimes{h_2}$ and we are considering that $\rho_h$ is density matrix for a pure state, it can easily be generalised for mixed states and $\kappa_i$ are eigenvalues of the density operator $\rho_h$. 'S' is the dimension of the square density matrix where E$\times$F=S, where E is dimension of $h_1$ and F is dimension of $h_2$. Also, it is known that this quantity is invariant under unitary similarity transformation.

\subsubsection{Partial Trace Hermitian case:}
Partial trace is a mathematical tool of describing individual properties of a composite system. Partial Trace of a density matrix, $\rho_h$ is denoted by $\rho_{h_1}=Tr_2(\rho_{h_{12}})$ and $\rho_{h_2}=Tr_1(\rho_{h_{12}})$, where $h=h_1\otimes{h_2}$. The prescription to find it is given as 
\begin{equation}
\rho_{h_1}=Tr_2(\rho_{h_{12}})=(\sum_{j''=1}^{F}{_2\bra{\xi_{j''}}\mathds{1}_2\rho_h\ket{\xi_{j''}}_2})=\sum_{(i,i'),j''=1}^{EF}c_{i',j''}c_{ij''}^*\ket{\xi_i'}_1{_1\bra{\xi_i}}\mathds{1}_1
\end{equation}
where $\ket{\xi_{j''}}_2\in\mathcal{H}_{h_2}$
Example: $\ket{\Xi}=\sqrt{\frac{1}{2}}\ket{0}_1\ket{1}_2+\ket{1}_1\ket{0}_2)$
here $$\ket{\xi_1}_1=\ket{0}_1 and \ket{\xi_2}_1=\ket{1}_1$$
$$\ket{\xi_1}_2=\ket{1}_2 and \ket{\xi_2}_2=\ket{0}_2$$
and $$c_{12}=c_{21}=0$$ and $$c_{11}=c_{22}=\sqrt\frac{1}{2}$$, E=F=2.
$$\rho_{h_1}=\frac{1}{2}\ket{0}_1{_1\bra{0}}+\frac{1}{2}\ket{1}_1{_1\bra{1}}$$ . Entanglement entropy of composite systems is defined as,
$$E(\Xi)=-Tr(\rho_{h_1}\log_2\rho_{h_1})$$ or another definition $$E(\Xi)=-Tr(\rho_{h_2}\log_2\rho_{h_2})$$

where $\ket{\Xi}$ is an entangled state and $\rho_{h}$ is the density operator for that state and, $\rho_{h_1}$ and $\rho_{h_2}$ are found by the prescription given above.

In explicit way they can be written as,
$$E(\Xi)=Tr((\sqrt{\mathds{1}_1}^{-1}\rho_{h_1}\sqrt{\mathds{1}_1})(\log_2\sqrt{\mathds{1}_1}^{-1}\rho_{h_1}\sqrt{\mathds{1}_1}))$$ This can be written due to the simple fact that $\sqrt{\mathds{1}}$ is always unitary and entropy is invariant under unitary transformations. It can also be equivalently given as, 
$$E(\Xi)=Tr((\sqrt{\mathds{1}_2}^{-1}\rho_{h_2}\sqrt{\mathds{1}_2})(\log_2\sqrt{\mathds{1}_2}^{-1}\rho_{h_2}\sqrt{\mathds{1}_2}))$$. Writing Entanglement entropy this way provides clarification later.

\section{Composite systems involving Pseudo-Hermitian Systems:}

\subsection{Case1 \texorpdfstring{$H=H_1 \otimes H_2$}{E=mc\texttwosuperior}}

Consider, case of a bipartite composite system of two Pseudo-Hermitian Hamiltonians given as $H_1\otimes{H_2}$
and having eigenvectors $\ket{\psi_i}_1$ and $\ket{\psi_j}_2$ s.t. $(\ket{\psi_i}_1\otimes\ket{\psi_j}_2)$ form a biorthogonal basis under metric $\eta=\eta_1\otimes{\eta_2}$
(Analogous to $\mathds{1}_1\otimes\mathds{1}_2$). $$\eta_1=\sum_{i=1}^{E}\ket{\phi_i}_1{_1\bra{\phi_i}}$$
and $$\eta_2=\sum_{j=1}^{F}\ket{\phi_j}_2{_2\bra{\phi_j}}$$
where $\ket{\phi_i}_1\in\mathcal{H}_{H_1^\dagger}$ and  $\ket{\phi_j}_2\in\mathcal{H}_{H_2^\dagger}$. Any state of Hilbert space of such a system is given as,
$$\ket{\Psi}=\sum_{i,j=1}^{EF}c_{ij}\ket{\psi_i}_1\otimes\ket{\psi_j}_2$$ where $\ket{\psi_i}_1\in\mathcal{H}_{H_1}$ and $\ket{\psi_j}_2\in\mathcal{H}_{H_2}$.
and  $\mathcal{H}_H=\mathcal{H}_{H_1}\otimes\mathcal{H}_{H_2}$ . $_1\bra{\psi_{i'}}\eta_1\ket{\psi_i}_1=\delta_{i'i}$ and  $_2\bra{\psi_{j'}}\eta_2\ket{\psi_j}_2=\delta_{j'j}$. Now, analogous to Hermitian case $\ket{\Psi}$ is called seperable if,
$\ket{\psi}=\sum_{i,j=1}^{EF}c_{ij}\ket{\psi_i}_1\otimes\ket{\psi_j}_2=(\sum_{i=1}^{E}c_i\ket{\psi_i}_1)\otimes(\sum_{j=1}^{F}c_j\ket{\psi_j}_2)$.
otherwise it is called entangled.
$$\bra{\Psi}\eta\ket{\Psi}=(\sum_{i,j=1}^{EF}c_{ij}^*{_1}\bra{\psi_{i}}\otimes{_2}\bra{\psi_{j}})(\eta_1\otimes{\eta_2})(\sum_{i',j'=1}^{EF}c_{i'j'}\ket{\psi_{i'}}_1\otimes\ket{\psi_{j'}}_2)$$ as we know,  $_1\bra{\psi_{i'}}\eta_1\ket{\psi_i}_1=\delta_{i'i}$ and  
$_2\bra{\psi_{j'}}\eta_2\ket{\psi_j}_2=\delta_{j'j}$. we have,
$$\bra{\Psi}\eta\ket{\Psi}=\sum_{i.j=1}^{EF}\mid{c_{ij}}\mid^2$$ which applies to both entangled and separable states.

\subsubsection{Density Matrix Formulation of case 1:}
Density matrix of such a composite system is given as $$\rho_H=\ket{\Psi}\bra{\Psi}(\eta_1\otimes\eta_2)$$ s.t.
$$Tr(\rho_H)=\bra{\Psi}(\eta_1\otimes\eta_2)\ket{\Psi}=\sum_{i.j=1}^{EF}\mid{c_{ij}}\mid^2$$ another important result is that as $$\eta=\eta_1\otimes\eta_2$$ and as we already saw, $\eta_1=\pi_1^\dagger\pi_1$ and $\eta_2=\pi_2^\dagger\pi_2$ then we have already seen that for a particular non composite system Hamiltonian, $\rho_H=\pi^{-1}\rho_h\pi$ and $\pi\ket{\Psi}=\ket{\Xi}$, where after transformation $\pi{H}\pi^{-1}=h$ we have $h=h^\dagger$. 
Now we check for composite system, that $\rho_H$ is equal to
$$(\sum_{i,j=1}^{EF}c_{ij}\ket{\psi_i}_1\otimes\ket{\psi_j}_2)(\sum_{i',j'=1}^{EF}c_{i'j'}^{*}{_2}\bra{\psi_{j'}}\otimes{_1}\bra{\psi_{i'}})(\eta_1\otimes{\eta_2})$$ 
as we know, $(A_1\otimes{B_2})(C_1\otimes{D_2})=A_1C_1\otimes{B_2D_2}$.
Then $\rho_H$ becomes 
$$(\pi_1^{-1}\otimes{\pi_2^{-1}})(\sum_{i,j=1}^{EF}c_{ij}\ket{\xi_i}_1\otimes\ket{\xi_j}_2)(\sum_{i'j'=1}^{EF}c_{i'j'}^*{_2}\bra{\xi_{i'}}\otimes{_1}\bra{\xi_{j'}})(\pi_1^{-1^\dagger}\otimes\pi_2^{-1^\dagger})(\pi_1{^\dagger}\pi_1\otimes\pi{_2^\dagger}\pi_2)$$
last two brackets become $$(\pi_1^{-1^\dagger}\otimes\pi_2^{-1^\dagger})(\pi_1^\dagger\pi_1\otimes\pi_2^\dagger\pi_2)=\pi_1\otimes\pi_2$$ 
defining $\pi_1\otimes\pi_2=\pi$ and $\pi_1^{-1}\otimes\pi_2^{-1}=\pi^{-1}$ we have 
$$\rho_H=\pi^{-1}\rho_h\pi$$. Hence, there exists a similarity transformation$$(\pi_1\otimes\pi_2)(H_1\otimes{H_2})(\pi_1^{-1}\otimes\pi_2^{-1})=\pi{H}\pi^{-1}=h $$ s.t. $h=h^\dagger$. Now, von neumann entropy as defined earlier gives $$V(\rho_H)=-Tr(\pi\rho_H\pi^{-1}\log_2(\pi\rho_H\pi^{-1}))$$ substituting $\rho_H=\pi^{-1}\rho_h\pi$ it gives $$V(\rho_H)=-Tr(\rho_h\log_2(\rho_h))=V(\rho_h)$$ 
Hence, we can see that given a composite system on Pseudo-Hermitian systems their entropies are same as that of composite system made of isospectral partners of individual Pseudo-Hermitian systems.
This can also be seen for Entanglement entropy. For that we first need to formalise the act of partial tracing for such a theory.

\subsubsection{Partial Tracing for Pseudo-Hermitian composite systems for case 1 :}
The density matrix of a system making a composite system is found for a Pseudo-Hermitian system by partial tracing as given below 
$$\rho_{H_1}=Tr(\rho_H)=\sum_{j''=1}^{F}{_2}\bra{\psi_{j''}}\eta_2\rho_H\ket{\psi_{j''}}_{2}$$
as seen earlier for hermitian case it can be reduced to,
$$\rho_{H_1}=\sum_{i,i',j''}^{EF}c_{i'j''}c_{ij''}^*\ket{\psi_{i'}}_1{_1}\bra{\psi_i}\eta_1$$ and  $$\rho_{H_2}=\sum_{i'',j,j'}^{EF}c_{i''j}c_{i''j'}^*\ket{\psi_{j'}}_2{_2}\bra{\psi_j}\eta_2$$

Using definition of Von Neumann entropy for single Pseudo-Hermitian systems as formalized earlier and defining Entanglement entropy for case 1 of composite systems containing these systems we see that, Entanglement entropy needs to be defined as 
$$E(\Psi)=-Tr(\pi_1\rho_{H_1}\pi_1^{-1}\log_2\pi_1\rho_{H_1}\pi_1^{-1})$$ as we have already seen that $\pi_1H_1\pi_1^{-1}=h_1$ and $\pi_1\ket{\psi_i}_1=\ket{\xi_i}_1$ we have $$\rho_{H_1}=\pi_1^{-1}\rho_{h_1}\pi_1$$ where $$\rho_{h_1}=\sum_{i,i',j''=1}^{EF}c_{i'j''}c_{ij''}^*\ket{\xi_{i'}}_1{_1}\bra{\xi_i}\mathds{1}_1$$
and therefore, Entanglement entropy $E(\Psi)$ becomes $$E(\Psi)=-Tr(\rho_{h_1}\log_2\rho_{h_1})=E(\Xi)$$
Hence we can see that the value of entanglement entropy of an entangled state from composite system of 2 Pseudo-Hermitian systems will be same as entangled entropy of an entangled state from a Hermitian isospectral system. All of these properties can be checked using a bipartite composite system 
\begin{equation}
H=\begin{pmatrix}
re^{i\theta} & s\\
s & re^{-i\theta}
\end{pmatrix}\otimes
\begin{pmatrix}
r'e^{i\theta'} & s'\\
s' & r'^{-i\theta'}
\end{pmatrix}
\end{equation}

\subsection{Case2 \texorpdfstring{$H=H_1\otimes{h_2}$}{E=mc\texttwosuperior}} 
Consider, the bipartite  composite system of a Pseudo-Hermitian ($H_1$) and a Hermitian ($h_2$) Hamiltonian systems. The eigenvectors of $H=H_1\otimes{h_2}$  will be $\ket{\psi_i}_1\otimes\ket{\xi_j}_2$ s.t. the, metric will be $\eta_1\otimes\mathds{1}_2$, where $\eta_1=\sum_{i=1}^{E}\ket{\phi_i}_1{_1}\bra{\phi_i}$ s.t. $\ket{\phi}\in\mathcal{H}_{H^\dagger}$. Any arbitrary state of such a composite system is given as, $$\ket{\Psi}=\sum_{i,j=2}^{EF}c_{ij}\ket{\psi_i}_1\otimes\ket{\xi_j}_2$$ where $\ket{\psi}_1\in\mathcal{H}_{H_1}$ and $\ket{\xi}_2\in\mathcal{H}_{h_2}$ and E and F are dimensions of $H_1$ and $h_2$.Also, $\mathcal{H_H}=\mathcal{H}_{H_1}\otimes{\mathcal{H}_{h_2}}$ s.t. $${_1}\bra{\psi_i}\eta_1\ket{\psi_{i'}}_1=\delta_{ii'}$$
and $${_2}\bra{\xi_j}\mathds{1}_2\ket{\xi_{j'}}_2=\delta_{jj'}$$
$\ket{\Psi}$ is called 'separable' if ,
,
$$\ket{\Psi}=(\sum_{i=1}^{E}c_{i}\ket{\psi_i}_1)\otimes(\sum_{j=1}^{F}c_{j}\ket{\xi_j}_2)$$ otherwise it is entangled.                             
Also, it can be seen that
\begin{equation}
\bra{\Psi}\eta\ket{\Psi}=(\sum_{i,j=1}^{E,F}c_{ij}^*\bra{\psi_i}_1\otimes\bra{\xi_j}_2)(\eta_1\otimes\mathds{1}_2)(\sum_{i',j'=1}^{E,F}c_{i'j'}\ket{\psi_{i'}}_1\otimes\ket{\xi_{j'}}_2)=\sum_{i,j=1}^{E,F}\mid{c_{i,j}}\mid^2
\end{equation} because, $_1\bra{\psi_i}\eta_1\ket{\psi_{i'}}_1=\delta_{ii'}$ and $_2\bra{\xi_j}\mathds{1}_2\ket{\xi_{j'}}_2=\delta_{jj'}$.

\subsubsection{Density Matrix Formulation for case 2:}
Density matrix of such a composite system is given as $$\rho_H=\ket{\Psi}\bra{\Psi}(\eta_1\otimes\mathds{1}_2)$$ s.t.
$$Tr(\rho_H)=\bra{\Psi}(\eta_1\otimes\mathds{1}_2)\ket{\Psi}=\sum_{i.j=1}^{EF}\mid{c_{ij}}\mid^2$$ another important result is that as $$\eta=\eta_1\otimes\mathds{1}_2$$ and as we already saw, $\eta_1=\pi_1^\dagger\pi_1$ and $\mathds{1}_2={\sqrt{\mathds{1}_2}}^\dagger{\sqrt{\mathds{1}}_2}$ then we have already seen that for a particular non composite system Hamiltonian, $\rho_H=\pi^{-1}\rho_h\pi$ and $\pi\ket{\Psi}=\ket{\Xi}$, where after transformation $\pi{H}\pi^{-1}=h$ we have $h=h^\dagger$. 
Now we check for composite system, we have $\rho_H$ is equal to
$$(\sum_{i,j=1}^{EF}c_{ij}\ket{\psi_i}_1\otimes\ket{\xi_j}_2)(\sum_{i',j'=1}^{EF}c_{i'j'}^{*}{_2}\bra{\xi_{j'}}\otimes{_1}\bra{\psi_{i'}})(\eta_1\otimes\mathds{1}_2)$$ which becomes after using previous similarity transform $\pi$
$$(\sum_{i,j=1}^{EF}c_{ij}(\pi_1^{-1}\ket{\psi_i}_1)\otimes(\sqrt{\mathds{1}}_2^{-1}\ket{\xi_j}_2))(\sum_{i',j'=1}^{EF}c_{i'j'}^{*}(_2\bra{\xi_{j'}}\sqrt{\mathds{1}}_2^{-1^\dagger})\otimes(_1\bra{\psi_{i'}}\pi_1^{-1^\dagger}))$$
as we know, $(A_1\otimes{B_2})(C_1\otimes{D_2})=A_1C_1\otimes{B_2D_2}$.
Then $\rho_H$ becomes 
$$(\pi_1^{-1}\otimes{\sqrt{\mathds{1}}_2^{-1}})(\sum_{i,j=1}^{EF}c_{ij}\ket{\xi_i}_1\otimes\ket{\xi_j}_2)(\sum_{i'j'=1}^{EF}c_{i'j'}^*{_2}\bra{\xi_{i'}}\otimes{_1}\bra{\xi_{i'}})(\pi_1^{-1^\dagger}\otimes\sqrt{\mathds{1}}_2^{-1^\dagger})(\pi_1^\dagger\pi_1\otimes\sqrt{\mathds{1}}_2^\dagger\sqrt{\mathds{1}}_2)$$
last two brackets become $$(\pi_1^{-1^\dagger}\otimes\sqrt{\mathds{1}}_2^{-1^\dagger})(\pi_1^\dagger\pi_1\otimes\sqrt{\mathds{1}}_2^\dagger\sqrt{\mathds{1}}_2)=\pi_1\otimes\mathds{1}_2$$ 
defining $\pi_1\otimes\sqrt{\mathds{1}_2}=\pi$ and $\pi_1^{-1}\otimes\sqrt{\mathds{1}_2}^{-1}=\pi^{-1}$ we have 
$$\rho_H=\pi^{-1}\rho_h\pi$$ where $h=h^{\dagger}$. Now, the von neumann entropy as defined earlier gives $$V(\rho_H)=-Tr(\pi\rho_H\pi^{-1}\log_2(\pi\rho_H\pi^{-1}))$$ substituting $\rho_H=\pi^{-1}\rho_h\pi$ it gives $$V(\rho_H)=-Tr(\rho_h\log_2(\rho_h))=V(\rho_h)$$ 
Hence, we can see that given a composite system of Pseudo-Hermitian  and a Hermitian system, their entropy values are same as that of composite system made of their isospectral hermitian partner of Pseudo-Hermitian systems embedded in it and of respective hermitian systems already present.
This can also be seen for Entanglement entropy. For that we again first need to formalize the act of partial tracing for such systems.

\subsubsection{Partial Tracing for Pseudo-Hermitian composite systems  for case 2:}
The density matrix of a Pseudo-Hermitian part of Composite case 2 system is given by partial tracing, below 
as seen earlier for hermitian case it can be reduced to,
$$\rho_{H_1}=\sum_{i,i',j''}^{EF}c_{i'j''}c_{ij''}^*\ket{\psi_{i'}}_1{_1}\bra{\psi_i}{\eta_1}$$ and  $$\rho_{H_2}=\sum_{i'',j,j'}^{EF}c_{i''j}c_{i''j'}^*\ket{\xi_{j'}}_2{_2}\bra{\xi_j}\mathds{1}_2$$

Using definition of Von Neumann entropy for single Pseudo-Hermitian systems as derived earlier and defining Entanglement entropy for case 2 of composite systems containing these individual systems we see that, Entanglement entropy needs to be defined as 
$$E(\Psi)=-Tr(\pi_1\rho_{H_1}\pi_1^{-1}\log_2\pi_1\rho_{H_1}\pi_1^{-1})$$  Using the case whichever maybe, also note that $\sqrt{\mathds{1}}$ is always unitary, instead it is equivalent definition of a unitary matrix. For first definition it can be seen that $\pi_1H_1\pi_1^{-1}=h_1$ and $\pi_1\ket{\psi_i}_1=\ket{\xi_i}_1$  and $h_1=h_1^\dagger$we have 
$$\rho_{H_1}=\pi_1^{-1}\rho_{h_1}\pi_1$$ 
where 
$$\rho_{h_1}=\sum_{i,i',j''=1}^{EF}c_{i'j''}c_{ij''}^*\ket{\xi_{i'}}_1{_1}\bra{\xi_i}\mathds{1}_1$$
and therefore, Entanglement entropy $E(\Psi)$ becomes $$E(\Psi)=-Tr(\rho_{h_1}\log_2\rho_{h_1})=E(\Xi)$$

Where $\ket{\Xi}$ in h is now a isospectral entangled state of $\ket{\Psi}$ , an entangled state in H. 
Hence we can see that the value of entanglement entropy of an entangled state from composite system of some Pseudo-Hermitian systems and some hermitian systems will be same as entangled entropy of an entangled state from a Hermitian isospectral system of the pseudo-hermitian and the already present hermitian systems. All of these properties can be checked using a bipartite composite system 

\begin{equation}
H=\begin{pmatrix}
re^{i\theta} & s\\
s & re^{-i\theta}
\end{pmatrix}\otimes
\begin{pmatrix}
r' & s'\\
s' & r'
\end{pmatrix}
\end{equation}

\section{Rate of entanglement for such composite systems :}
It can now be seen that rate of entanglement of composite systems having Pseudo-Hermitian systems will have same value as that of an equivalent Hermitian isospectral Composite system. As per the definition of Rate of entanglement production by a particular composite system H for a particular entangled state $\Psi$ we write $$\Gamma_H=\frac{dE(\Psi)}{dt}$$  As we already saw,Entanglement entropies $E(\Psi)=E(\Xi)$ where $\Psi$ is an entangled state which belongs to Pseudo-Hermitian composite system H and $\Xi$ belongs to it's isospectral composite hermitian system h. We have,
$$\Gamma_H=\frac{dE(\Psi)}{dt}=\Gamma_h=\frac{dE(\Xi)}{dt}$$

\section{Conclusion:}
We learned that Theory of Pseudo-Hermiticity can be consistently applied to dissipative quantum systems ,\cite{Dekker1975,Dekker1981,MorseFeshbach,Sergi}. It was presented in Density Operator language, which has it's own pros and cons. A few clarifications on ,\cite{AKPati, Wang}were provided. We saw that the Von-Neumann entropy and Entanglement Entropy of these Non-Hermitian systems (in real eigenvalue regime)are same a their Hermitian counterparts. Hence, it does not have effect on the rate of Entanglement creation as shown in \cite{AKPati, Wang}, nor can composite systems of these systems be used to distinguish the Non-Hermitian and their Hermitian iso-spectral partners.

\section*{Appendix}
Here we will give examples of the results obtained earlier. Firstly we give the example of a $2\times2$ Non-Hermitian matrix in the real eigenvalue regime, and show how it is related to an isospectral-hermitian hamiltonian, both algebraically and geometrically. Next we will see by analogy the geometry of a composite Pseudo-Hermitian system. The following Appendix is based on ,\cite{Mosseri,Brody,Beng,AnandanAharanov}.

Consider a $2\times2$ Hermitian Hamiltonian as given below

\begin{equation}
\begin{pmatrix}
s & r\\
r & s
\end{pmatrix}
\end{equation}

it's eigenvectors are 
$$
\sqrt{\frac{1}{2}}\begin{pmatrix}
1 \\
1 
\end{pmatrix}
and
\sqrt{\frac{1}{2}}\begin{pmatrix}
1 \\
-1 
\end{pmatrix}
$$ and eigenvalues are real hamiltonian is hermitian.

When (41) is unitarily transformed it becomes 
\begin{equation}
\begin{pmatrix}
s & re^{-i\theta}\\
re^{i\theta} & s
\end{pmatrix}
\end{equation}

which has eigenvectors
$$
\sqrt{\frac{1}{2}}\begin{pmatrix}
e^{i{\theta/2}} \\
e^{-i{\theta/2}} 
\end{pmatrix}
and
\sqrt{\frac{1}{2}}\begin{pmatrix}
e^{i{\theta/2}} \\
-e^{-i{\theta/2}}
\end{pmatrix}
$$

We can see in both cases eigenvectors are orthogonal to each other. As we know the state space of a Hermitian $2\times2$ system can be projected onto a Bloch Sphere, always. we can see that geometrically the act of unitary transformation just rotates the Bloch sphere. It can be seen as follows, by calculating the Bloch vector, it shows that for the eigenvector of Hamiltonian in eq(41) 
$$\bar{S}=1\hat{x}$$ and for eq(42) it is $$\bar{S}=cos(\theta)\hat{x}+sin(\theta)\hat{y}$$ which shows that their Bloch vectors are only rotated by $\theta$ angle. It can also be seen that the geodesic distance between both eigenvectors of eq(41) and of eq(42) have distance $\pi$ and hence they are orthogonal to each other. The geodesic distance between $\xi_i$ and $\xi_j$ is given as 
$$\delta=arccos(\sqrt{\frac{(\xi_i^\dagger\xi_j)(\xi_j^\dagger\xi_i)}{(\xi_i^\dagger\xi_i)(\xi_j^\dagger\xi_j)}})$$. Moreover, the Eigenvalues and expectation values of the observables of these systems will be real as is already known.
However, the Hamiltonian

\begin{equation}
H=\begin{pmatrix}
re^{i\theta} & s\\
s & re^{-i\theta}
\end{pmatrix}
\end{equation}
is not Unitarily equivalent to eq(41) or eq(42) Hamiltonian. Instead, if a matrix is untarily equivalent to one matrix then it is unitarily equivalent to all other matrices which are unitarily equivalent to the first matrix. This, Hamiltonian is Non-Hermitian with eigenvalues  $$\lambda_{\pm}=rcos(\theta)\pm\sqrt{{s^2}-{r^2}{sin^2}(\theta)}$$
The eigenvectors for $\lambda_{+}$ and $\lambda_{-}$ in the real eigenvalue regime (i.e. when 

${s^2}-{r^2}{sin^2}(\theta)>0 )$ respectively, are

$$
\psi\pm=\sqrt{\frac{1}{2cos(\alpha)}}\begin{pmatrix}
e^{i\alpha/2} \\
e^{-i\alpha/2}
\end{pmatrix}
and
\sqrt{\frac{1}{2cos(\alpha)}}\begin{pmatrix}
e^{-i\alpha/2} \\
-e^{i\alpha/2} 
\end{pmatrix}
$$
where, $sin\alpha=\frac{r}{s}sin{\theta}$ and eigenvalues become $\lambda_\pm=rcos\theta\pm{scos\alpha}$. It can be easily seen that the eigenvectors are not orthonormal to each other and if we find the eigenvectors of $H^\dagger$ in the real eigenvalue regime then they turn out to be. 

$$
\phi\pm=\sqrt{\frac{1}{2cos(\alpha)}}\begin{pmatrix}
e^{-i\alpha/2} \\
e^{i\alpha/2}
\end{pmatrix}
and
\sqrt{\frac{1}{2cos(\alpha)}}\begin{pmatrix}
-e^{i\alpha/2} \\
e^{-i\alpha/2} 
\end{pmatrix}
$$

It can be seen that $\psi_{\pm}^\dagger\phi_{\pm}=1$ and $\psi_{\mp}^\dagger\phi_{\pm}=0$, we can create an operator $\eta=\phi_+\phi_+{^\dagger}+\phi_-\phi_-^{\dagger}$ s.t we get all the eigenvectors of eq(43) be orthogonal to each other i.e. $\psi_{\pm}^\dagger(\eta\psi_{\pm})=1$ and $\psi_{\mp}^\dagger(\eta\psi_{\pm})=0$
where, now representation of $\eta$ =

\begin{equation}
\frac{1}{cos\alpha}\begin{pmatrix}
1 & -isin{\alpha}\\
isin{\alpha} & 1
\end{pmatrix}
\end{equation}

and square root of $\eta$ which is found by diagonalizing $\eta$ and then taking square roots of the diagonal elements of $\eta$ and then again transforming the root diagonal form to root non-diagonal form. We, get
\begin{equation}
\pi=\frac{1}{2}\begin{pmatrix}
(a+b)& -i(a-b)\\
i(a+b) & (a+b)
\end{pmatrix}
\end{equation}
Where $a=\sqrt{sec\alpha+tan\alpha}$ and $b=\sqrt{sec\alpha-tan\alpha}$ it can be easily be seen that $\pi$ is not unitary ($\pi^\dagger\neq\pi^{-1}$).
\begin{equation}
\pi^{-1}=\frac{1}{2}\begin{pmatrix}
(a+b)& i(a-b)\\
-i(a+b) & (a+b)
\end{pmatrix}
\end{equation}
It can now be checked that H can be transformed to an isospectral Hermitian hamiltonian by a similarity transform using $\pi$ s.t. $\pi{H}\pi^{-1}=h$ and $h=h^\dagger$. For, this case the Hamiltonian turns out to be 
\begin{equation}
h=\begin{pmatrix}
rcos\theta & scos\alpha\\
scos\alpha & rcos\theta
\end{pmatrix}
\end{equation} with eigenvectors,
\begin{equation}
\xi_+=\sqrt{\frac{1}{2}}\begin{pmatrix}
1\\
1
\end{pmatrix}
and
\xi_-=\sqrt{\frac{1}{2}}\begin{pmatrix}
1\\
-1
\end{pmatrix}
\end{equation}

It can be easily seen that it's eigenvalue is same as H, the eigenvectors $\xi$ being, $\pi\psi_{\pm}=\xi_{\pm}$ s.t. $\psi_{\pm}^\dagger\eta\psi_{\pm}=\xi_{\pm}^\dagger\pi^{-1^\dagger}\eta\pi^{-1}\xi_{\pm}=\xi_\pm^\dagger\xi_\pm=1$ and $\xi_\mp^\dagger\xi_\pm=0$. Hence the Geodesic distance between $\psi_\mp$ and $\psi_\pm$ is $\pi$ as their values are same as the orthogonal eigenvectors of h and it is well known that the geodesic distance between orthogonal vectors under metric $\mathds{1}$ is $\pi$, where $\pi$ here is angle in radians. This can also be checked by using the formula for geodesic distance between two states in the projective space of the state space of a Pseudo-Hermitian system, given by, 
$$\delta=arccos(\sqrt{\frac{(\psi_i^\dagger\eta\psi_j)(\psi_j^\dagger\eta\psi_i)}{(\psi_i^\dagger\psi_i)(\psi_j^\dagger\psi_j)}})$$ Check ,\cite{AnandanAharanov}.

Note now if we try to find the Bloch Vectors of corresponding eigenvectors of eq(43) using usual Pauli matrices then $\psi_+^\dagger\eta\sigma_z\psi_+$ gives value $itan(\alpha)$ which is a complex value. Hence, the expectation value of spin will be complex, so the observables need to be changed. We will see that indeed we just changed the representation of the observables however they will give same expectation values for corresponding isospectral states.
As, we already saw the observable given  by Hamiltonian is changed as $\pi{H}\pi^{-1}=h$ the Pauli matrices will also change as $\pi^{-1}\bar{\sigma}\pi=\bar{\sigma}_{new}$ s.t. now  it can be easily seen that $S_{\psi_+}=\hat{x}$ and hence the projective space of H is same as that of h, which is Hermitian. We can check that only the arrangement of the state space has changed between the Pseudo-Hermitian and iso-spectral Hermtian system. All the expectation values will match between the systems. All of the above properties can be visualized as given by Fig.1.

\begin{figure}[h!]
	\includegraphics[scale=0.15]{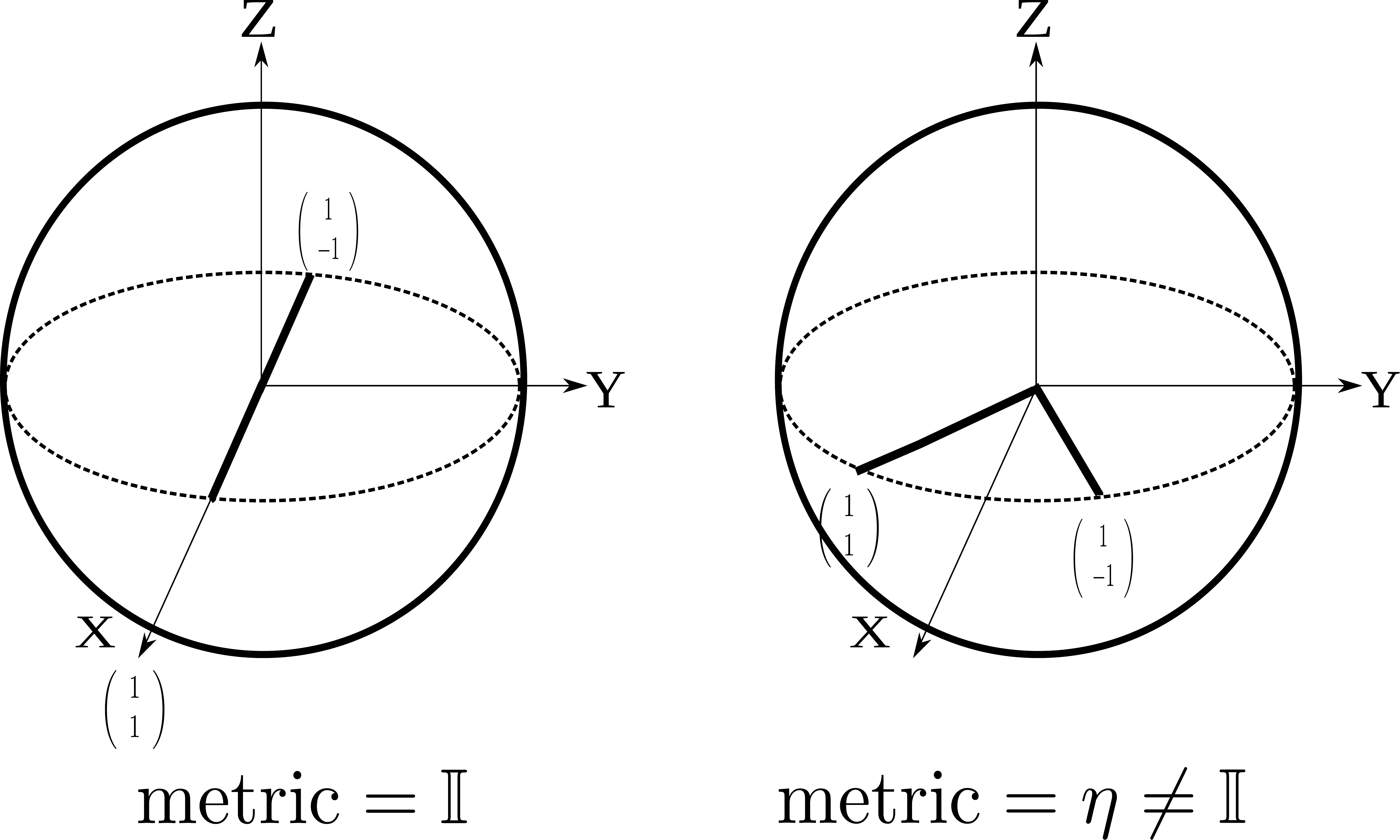}
	\caption{The dark lines represent $\sqrt{\frac{1}{2}}(\ket{0}\pm\ket{1})$, but note they are not equivalent to each other after similarity transformation with $\pi$}
\end{figure}

Using all this it can be easily visualised that for a composite system, for example consider 
\begin{equation}
H=\begin{pmatrix}
re^{i\theta} & s\\
s & re^{-i\theta}
\end{pmatrix}\otimes\begin{pmatrix}
r'e^{i\theta'} & s'\\
s' & r'e^{-i\theta'}
\end{pmatrix}
\end{equation}

We will have for a separable state of such a composite system that the Complex projective space of it can be decomposed into two individual Bloch spheres representing the two states making up this seperable state. Also, an Entangled state will be represented by an $S^7$( Unit sphere in 8 dimensions) as is done in Hermitian case.


\begin{thebibliography}{9}

\bibitem{Bender1}
Bender, Carl M. and Brody, Dorje C. and Jones, Hugh F., \emph{Must a Hamiltonian be Hermitian?}, American Journal of Physics, Vol. 71, No. 11, 1095-1102, (2003).

\bibitem{Bender2}
Bender, Carl M. and Boettcher, Stefan, \emph{Real Spectra in Non-Hermitian Hamiltonians Having PT-Symmetry}, Phys. Rev. Lett., Vol. 80, issue. 24, 5243-5246 (1998).
	

\bibitem{Bender3}
Bender, Carl M. and Boettcher, Stefan and Meisinger, Peter N., \emph{PT-symmetric quantum mechanics}, Journal of Mathematical Physics, Vol.  40, 2201-2229, (1999).	

\bibitem{Mostafazadeh}
Ali Mostafazadeh, \emph{PSEUDO-HERMITIAN REPRESENTATION OF QUANTUM MECHANICS}, International Journal of Geometric Methods in Modern Physics, Vol. 7, 1191-1306, (2010).	

\bibitem{Dekker1975}
Dekker H.,\emph{On the quantization of dissipative systems in the Lagrange-Hamilton formalism}, Zeitschrift für Physik B Condensed Matter, Vol. 21, 295-300, (1975).



\bibitem{Dekker1981}
Dekker H.,\emph{Classical and quantum mechanics of the damped harmonic oscillator}, Physics Reports, Vol. 80, 1-110, (1981).

\bibitem{MorseFeshbach}
Morse, P.M.C. and Feshbach, H., \emph{Methods of theoretical physics}, pg. 298, (1953).
	

\bibitem{Guo}
Guo, A. and Salamo, G. J. and Duchesne, D. and Morandotti, R. and Volatier-Ravat, M. and Aimez, V. and Siviloglou, G. A. and Christodoulides, D. N., \emph{Observation of PT-Symmetry Breaking in Complex Optical Potentials},Phys. Rev. Lett., Vol. 103, (2009).
	

\bibitem{AKPati}
A. K. Pati, \emph{Entanglement in non-Hermitian quantum theory}, Pramana, Vol. 73, 485-498 (2009).
	

\bibitem{AnandanAharanov}
Anandan, J. and Aharonov, Y., \emph{Geometry of quantum evolution}, Phys. Rev. Lett., Vol. 65, 1697--1700, (1990).
	

\bibitem{Beng}
Bengtsson, I. and {\.Z}yczkowski, K., \emph{Geometry of Quantum States: An Introduction to Quantum Entanglement}, (2006).


\bibitem{Brody}
Dorje C. Brody and Lane P. Hughston, \emph{Geometric quantum mechanics}, Journal of Geometry and Physics, Vol. 38, 19-53, (2001).

\bibitem{Sakurai}
J. Jun John Sakurai and Napolitano, J., \emph{Modern Quantum Mechanics}, (2010).
	
\bibitem{Lang}
Lang, S, \emph{Linear Algebra}, (2010).
	

\bibitem{McMahon}
McMahon, D., \emph{Quantum Computing Explained}. (2007).

\bibitem{Mosseri}
Mosseri, R. and Dandoloff, R., \emph{Geometry of entangled states, Bloch spheres and Hopf fibrations}, (2001).


\bibitem{Sergi}
{Sergi}, A. and {Zloshchastiev}, K.~G., \emph{Non-Hermitian Quantum Dynamics of a Two-Level System and Models of Dissipative Environments}, International Journal of Modern Physics B, Vol. 27, (2013).
	


\bibitem{Japaridze}
G S Japaridze, \emph{Space of state vectors in  PT-symmetric quantum mechanics}, Journal of Physics A: Mathematical and General, Vol. 35, 1709, (2002).
	

\bibitem{Bender4}
Bender, Carl M. and Brody, Dorje C. and Jones, Hugh F., \emph{Complex Extension of Quantum Mechanics}, Phys. Rev. Lett., Vol. 89, (2002).


\bibitem{Wang}
Zielinski, Christian and Wang, Qing-hai, \emph{Entanglement Efficiencies in PT-Symmetric Quantum Mechanics}, International Journal of Theoretical Physics, (2012).
	










\end{thebibliography}
\end{document}